\documentstyle[aps,prl,epsf]{revtex}
\begin{document}
\twocolumn[\hsize\textwidth\columnwidth\hsize\csname
@twocolumnfalse\endcsname

\title{Surface defreezing of glasses}
\author{E. A. Jagla$^1$ and E. Tosatti$^{1,2,3}$}
\address{$^{(1)}$The Abdus Salam International Centre for Theoretical
Physics (ICTP), I-34014 Trieste, Italy}
\address{$^{(2)}$International School for Advanced Studies (SISSA-ISAS),
I-34014 Trieste, Italy}
\address{$^{(3)}$Istituto Nazionale per la Fisica della Materia (INFM),
Unita' SISSA, Trieste, Italy}
\maketitle

\begin{abstract}
A glass surface may still flow below the bulk glass transition temperature, 
where the underlying bulk is frozen. Assuming the existence at $T=T_0$ of
a bulk 
thermodynamical glass transition, we show that the 
glass-vapor interface is generally wetted by a liquid layer of 
thickness $l\sim -\ln(T_0-T)$ when $T\rightarrow T_0^-$. 
Contrary to standard surface melting of crystals however, 
the integrated value of the diffusivity across the interface remains
finite for $T\rightarrow T_0^-$. 
Difference in shape induced by bulk and by surface flow is discussed
as a possible means of experimental detection of surface defreezing.
\end{abstract}
 
\vskip2pc] \narrowtext

Glasses embody the paradigm of broken ergodicity in condensed classical
systems. While most approaches to glasses address an infinite, homogeneous 
system, there is a clear scope for extension to inhomogeneous situations,
as they may be variously realized in real life. The most 
common --and conceptually the simplest-- kind of inhomogeneity is 
represented by the surface. The problem we wish to address here is the 
state of a free glass surface, in the neighborhood of the bulk 
glass transition temperature $T_0$.

In crystals near the bulk melting temperature $T_M$, surface melting is well 
documented both experimentally and theoretically. As $T\rightarrow T_M$
along the solid-vapor bulk coexistence line, most crystalline faces of 
a majority of substances 
develop a microscopic liquid film which spontaneously wets, in full thermal 
equilibrium, the solid-vapor interface\cite{surfmelting1,surfmelting2}. 
The thickness of the film 
diverges as $T\rightarrow T_M$, explaining among other things
why crystals with free surfaces cannot sustain overheating. The 
surface acts as a `defect' where the liquid phase microscopically nucleates. 

Due in part to the lack of a comparable
understanding of the properties of even bulk glasses, the possibility
that similar phenomena might take place at the surface of glassy materials 
has been paid very little attention so far, in spite of its potential 
importance, both conceptual and practical. Conceptually, the purely 
dynamical arrest typical of glasses provides a neater example of broken
ergodicity than that of the solid, where the crystalline order parameter
represents an additional complication. Practically, the possible 
flow of surfaces can 
be expected to play a role in measurable properties of glasses, such as
friction, surface flow under intense acceleration, interfacial contact,
and other phenomena. 
In this letter, we base on a thermodynamic glass transition theory, of the
type recently considered within the glass community\cite{speedy}, 
our first attack to the glass surface problem.  

Given a thermodynamic formulation with a well
defined free energy versus some order parameter, such as the atomic
density $n(r)$, or the energy density $e(r)$, the natural approach to 
try first is a Landau theory\cite{surfmelting2}. 
For crystals, it provides  what certainly 
is the simplest microscopic theory of surface melting, summarized as
follows. Call $f(n)$ the bulk free energy density,
a function of the (uniform) atomic density $n$. The global free energy cost 
of a solid-vapor interface from the solid at $x =-\infty$, to the vapor at
$ x = + \infty$ is

\begin{equation}
F[n(x)] =  \int_{-\infty}^{+\infty} \left [ f(n(x)) + (J/2) 
(dn(x)/dx)^2 \right ] dx
\label{freeenergy}
\end{equation}
Strictly short range forces are assumed, and the gradient term ($ J > 0$) 
accounts as usual for reluctance of the order parameter $n$ against spatial 
change, assumed to be sufficiently slow. 
Along the solid-vapor coexistence line, and just below the triple point $T_M$,
$f(n)$ will exhibit, besides the two identically deep (solid and vapor) 
minima $f(n^s) = f(n^v)$ at $n=n^s$ and $n^v$ respectively, a third, 
shallower minimum $f(n^l)=  f(n^s) + \Delta$ at the intermediate liquid 
density $n^l$, in which $\Delta = \varepsilon_0 (T_M - T)$, where
$\varepsilon_0 > 0 $ is a constant.

Minimization of the free energy (\ref{freeenergy}) is formally 
identical, via the
dictionary $n \rightarrow  z$, $x \rightarrow t$, 
to minimization of the action of a classical point particle of coordinate
$z$ and mass $J$, moving in a potential $V(z) = - f(z)$,
with energy exactly equal to $-f(n^s) =- f(n^v)$. 
The resulting equilibrium interface profile $n(x)$
exhibits, between the solid and the vapor, an intermediate liquid film.
Calling $\gamma$ the curvature of $f(n)$ at the liquid minimum, the
thickness of the liquid film is given by $l \simeq - \sqrt{J/\gamma} \ln
(\Delta)$, which diverges as
$T$ approaches $T_M$ from below,
demonstrating surface melting (in real crystals, the logarithmic 
divergence
usually turns to a power law, due to long range forces, not included in 
(\ref{freeenergy})\cite{surfmelting1,surfmelting2}).

To apply arguments similar to these to a glass-vapor
interface we must have, as a starting point, a thermodynamical description for 
the properties of bulk glasses. For that, a first 
necessary assumption is that crystalline states of the system,
even if lower in free energy than any of the glass configurations,
can be ignored.
Experimentally, crystallization
of glasses can be kinetically avoided by a sufficiently 
rapid cooling in absence of crystalline germs. With this
assumption we can consistently speak of the glass
transition as an equilibrium phenomenon, even though it actually occurs on a
metastable branch of the phase diagram.

We will thus suppose that there is a thermodynamic glass
transition for the bulk system.
The configurational entropy of 
the system $s_c$ as a function of the 
enthalpy $h$ (to be used instead of the internal energy $e$ since we
work at constant pressure) should vanish for $h$ lower than some $h_0$, and
increase linearly for $h$ greater than $h_0$. The glass transition is in
this scheme a second order, mean field transition \cite{someone}. 
The critical temperature
$T_0$ is given by $T_0^{-1}=\left .\partial s_c/ \partial h \right
|_{h_0}$. For $T<T_0$ the system freezes in the  configurational 
ground state, and thermodynamical variables such as specific heat give
information only about the vibrational structure of the valley 
around the ground state.
For $T>T_0$  a configurational contribution arises, 
that counts the number of different valleys that the system is able to sample. 
The specific heat has a finite jump at
$T=T_0$, sometimes referred to as the Kauzman temperature\cite{vogel,nota}.

A further connection of this thermodynamic picture with dynamical
properties {\em at equilibrium} is given by the phenomenological 
Adam-Gibbs formula\cite{adam65}, that relates the
configurational entropy to dynamical variables such as the viscosity,
or the diffusivity $D$, in the form
$D=D_0 \exp\left ( -\frac{A}{Ts_c} \right )$
where $D_0>0$ and $A>0$ are constants. 
This formula can be made heuristically plausible\cite{adam65} but is
not rigorous, and should be considered just a useful working
hypothesis.
As $T\rightarrow T_0$,  $s_c$ and  thus $D$ vanish, reflecting the
impossibility for the system to jump between valleys, 
for only one is thermodynamically favored.
When applied to the previous description of a bulk glassy phase,
the Adam-Gibbs formula is consistent with a Vogel-Fulcher-Tammann
type temperature dependence of the diffusivity, which is 
experimentally well verified for many different glass formers\cite{vogel}. 

Next, we will assume our system to possess in addition to the glass-liquid
transition, 
a liquid-vapor transition. At each temperature below the liquid-vapor critical
temperature the free energy $f(h)$ will have two minima, one corresponding to
the vapor and the other to the condensed phase, the same minimum comprising
both liquid and glass because, as Fig. \ref{f2} indicates,
the glass-liquid
transition is second order. At the very minimum the 
system will be liquid or glassy 
depending on whether $T$ is lower or larger than $T_0$. 
The second order nature of
the glass-liquid transition reflects in a jump in the 
second derivative of $f(h)$ at
some critical value of the enthalpy $h_c$ ($h_c$ 
may be different from $h_0$ due to
vibrational contributions). 
On the phase boundary between the vapor 
and the dense phase the two minima of the free energy 
are degenerate, $f(h_1)=f(h_2)$.

We can now model the glass-vapor interface at bulk
coexistence, where the bulk enthalpies of the two coexisting
phases are $h_1$ and $h_2$. We will suppose, as in the
crystal-vapor interface, that the system can be
assigned a well defined, slowly varying enthalpy density 
$h(x)$ at each position $x$
across the interface. Moreover, for $x\rightarrow -\infty$ ($+\infty$), 
$h\rightarrow
h_1$ ($h_2$), forcing the existence of the interface through
the boundary conditions. In the end, we will
derive the interface profile $h(x)$ by minimizing a 
free energy functional qualitatively similar to that in
(\ref{freeenergy}), but using enthalpy as the order parameter
instead of density, and with only
two minima instead of three.

\begin{figure}
\narrowtext
\epsfxsize=3.3truein
\vbox{\hskip 0.05truein
\epsffile{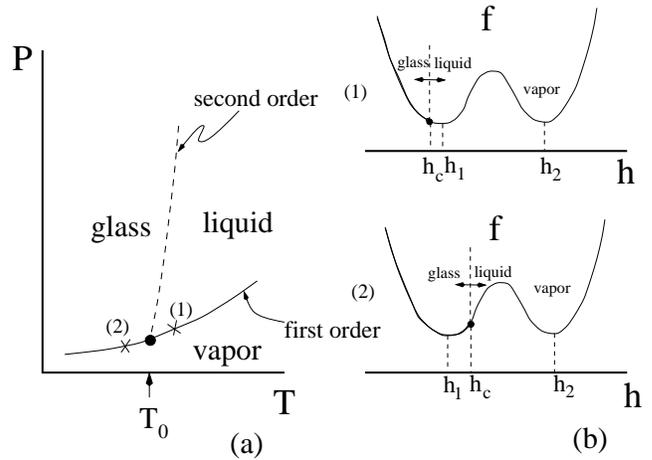}}
\medskip
\caption{(a) Generic pressure-temperature phase diagram for a system with
glass, liquid, and vapor phases (a possible crystalline phase is not
indicated). (b) Free energy density as a function of the enthalpy at two
coexistence points on the first order line of (a). The coexistence is
between liquid and vapor in (1), and between glass and vapor in (2).}
\label{f2}
\end{figure}

\begin{figure}
\narrowtext
\epsfxsize=3.3truein
\vbox{\hskip 0.05truein
\epsffile{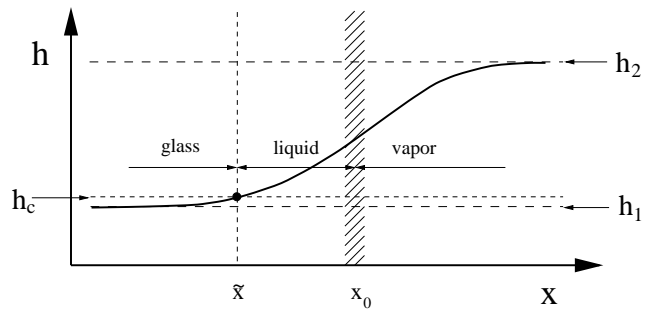}}
\medskip
\caption{Evolution of the interface enthalpy density
from the glass at the left to the vapor
at the right ($T<T_0)$. The glass-vapor interface is coated by a liquid film
of thickness $l \sim x_0-\tilde x$.}
\label{f3}
\end{figure}

A schematic representation of the interface enthalpy profile
$h(x)$  in the case $T<T_0$ is
given in Fig. \ref{f3}. This function interpolates between the two minima
$h_1$ and $h_2$ of Fig. \ref{f2}(b). We expect a weak singularity of $h(x)$
where $h=h_c$, the point in which $f(h)$ has a jump in the second
derivative. The corresponding point $\tilde x$ marks the border between
the glass at the left, an the liquid at the right. The liquid transforms
into the vapor phase roughly at some $x=x_0$, where the maximum of $f(h)$ is
overcome, and the rate of change of $h(x)$ should be maximum. 
The difference $ l = x_0-\tilde x$ can be identified with
the thickness of the liquid layer that wets the glass-vapor interface.
We are interested in the behavior of the system
around $T=T_0$. Exactly at this point $h_c$ coincides with the minimum of $f$,
and for slightly different values it may be supposed to have  a linear
dependence on $T_0-T$, namely $h_c-h_1=\varepsilon(T_0-T)$.


To calculate $h(x)$ and from that $l$, we proceed in the following way. 
As already mentioned,  
the free energy $f(h)$ is smooth except in the point $h_c$, where it has
a jump in the second derivative. We will approximate $f(h)$ with its quadratic
expansion close to $h_c$

\begin{eqnarray}
f(h)&=&\gamma' h^2/2~~~~~~~~~~~~~~~~~~~~~~~~~~~~~~~~~~ h<h_c\nonumber\\
f(h)&=&\gamma(h-h_c(1-\gamma'/\gamma))^2/2+f_0 ~~~~~h>h_c
\label{fdeh}
\end{eqnarray}
where we have taken $h_1=0$ for simplicity. $\gamma$ and $\gamma '$ (
$\gamma>\gamma'$) are
constants related to the bulk properties of the material, and  $f_0$ is a
constant ensuring continuity of $f$ at $h_c$.
The Euler equation satisfied by the equilibrium profile $h(x)$ is, 
from (\ref{freeenergy}) 
\begin{equation}
df(h)/dh= J d^2 h/dx^2
\label{eq}
\end{equation}
For $f(h)$ given by (\ref{fdeh}), the solution can be written in term of
exponentials, and a direct calculation shows that
\begin{eqnarray}
h(x)&=&A e^{\sqrt{\gamma'/J} x}   ~~~~ ~~~~~~~~~~~~~~~~~ h<h_c\nonumber\\
h(x)&=& u+(h_m-u)e^{\sqrt{\gamma/J} x}    ~~~~~~ h>h_c
\label{hs}
\end{eqnarray}
where $u=h_c(1-\gamma'/\gamma)$ and
$A=h_c\left[(h_m-u)/(h_c-u)\right]^{\sqrt{\gamma'/\gamma}}$
are values that depend on temperature through $h_c$.
Here we have set $x_0 = 0$, assuming that the barrier of 
$f(h)$ is parabolic up to
the maximum, that occurs precisely at $h_m$. 
Then the solution (\ref{hs}) describes the interface in the condensed region.
Deviations are expected to occur around the liquid-vapor transition.
The position $\tilde x$ of the liquid-glass interface,
namely the point at which $h(\tilde x)=h_c$ can be written as

\begin{equation}
\tilde x=-\sqrt{J/\gamma}\ln\left[(h_m-h_c(1-\gamma'/\gamma))
/(h_c\gamma'/\gamma)\right]
\end{equation}
This value diverges when $T\rightarrow T_0^-$, indicating the presence of a 
diverging liquid layer of thickness

\begin{equation}
\l \simeq -\sqrt{J/\gamma}\ln\left[\frac{\varepsilon(T_0-T)\gamma'}{h_m\gamma} \right].
\label{divergencia}
\end{equation}
The spatial scale for the thickness of the liquid layer is thus given by
$\sqrt{J/\gamma}$ , exactly as for a liquid layer at
the crystal-vapor interface.
This underlines a common physical origin, namely 
a lower surface free energy of the liquid 
relative to the solid. 

The logarithmic increase of the liquid layer thickness close to $T_0$
is at first sight also the same as that of the liquid layer in short range
crystal-vapor surface melting. However the conceptual origin of the logarithmic
behavior is different here, as it relates to a second order liquid-solid
transition, as opposed to a first order one in the crystal case.
In practice too, glass surface melting should be quite different. 
In crystals, the density
jump associated with melting gives inevitably rise to 
different optical conductivities of solid and liquid, leading to a nonzero 
Hamaker constant $H$ and to long range dispersion forces $\sim H/l^2$
\cite{hamaker}.
A positive Hamaker constant
will thus enhance surface melting, and generally transform the 
logarithmic into a power law growth, $l \sim (T_M-T)^{-1/3}$
\cite{surfmelting1,surfmelting2}.
A negative Hamaker constant will instead suppress, or block, surface
melting \cite{trayanov}. So will, for a different reason, 
commensuration of surface layering 
with the spacing of crystal planes\cite{layering}. 
In the glass there are no crystal planes to block surface melting. 
Moreover, the bulk density and optical conductivity of glass and liquid 
are not expected to differ discontinuously and 
the Hamaker constant should basically vanish. In conclusion, in
glass surface melting the logarithmic film growth behavior should be more
robust. 

\begin{figure}
\narrowtext
\epsfxsize=3.3truein
\vbox{\hskip 0.05truein
\epsffile{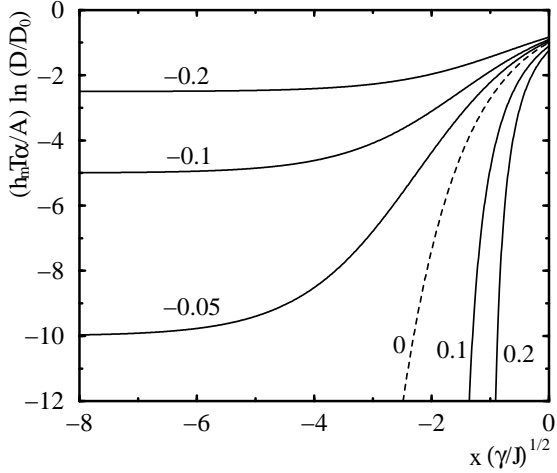}}
\medskip
\caption{Diffusivity across the glass-vapor interface for different values
of $\Delta/h_m = \varepsilon_0 (T_M - T)/h_m$, indicated on the curves 
(the particular value $\gamma'/\gamma =0.5$
was used). The dense phase (glass for $\Delta>0$, liquid for $\Delta<0$)
is at the left, while for $x {\protect \gtrsim}  0$ the
system is in the vapor phase. The curve for the bulk
glass transition temperature $T=T_0$ is shown as a dashed line.
Note the finite diffusivity in a ``melted'' surface film for 
$T < T_0$, and also the residually higher surface diffusivity
just above $T=T_0$.}
\label{f4}
\end{figure}

We can also calculate the diffusivity
profile $D(x)$. The configurational 
entropy vanishes linearly with  $h$ close to the liquid-glass transition, 
namely
\begin{eqnarray}
s_c(x)=
&{\alpha}&(h(x)-h_c)=\nonumber\\ 
&{\alpha}&\left[e^{\sqrt{\gamma
/J}x}\left[h_m-h_c\left(1-\frac{\gamma'}{\gamma}\right)
\right]-h_c\frac{\gamma'}{\gamma}\right],
\end{eqnarray}
where $\alpha$ is some constant. Using the Adam-Gibbs formula, we obtain

\begin{eqnarray}
\ln \left ( D(x)/D_0\right )  =
\frac{A}{T\alpha}\times\nonumber\\
\times \left[ \varepsilon (T_0-T) \left[ 
\frac{\gamma'}{\gamma}+(1-\frac{\gamma'}{\gamma})
e^{\sqrt{\gamma/J}x} 
\right] -h_m e^{\sqrt{\gamma/J}x}\right ]^{-1},
\label{d}
\end{eqnarray}
valid for $T$ close to $T_0$, and $x$ close to the liquid-glass
transition value $\tilde x$. For larger values of $x-\tilde x$, the
non-quadratic nature of the free energy functional (particularly the maximum
between the vapor and condensed phases) should be taken into account properly. 
If $T<T_0$, the previous expression is valid for $x>\tilde x$, and $D$ is
0 for $x<\tilde x$. If $T>T_0$, i.e., if the bulk system is in the liquid 
phase, then (\ref{d}) is valid for all $x$. In
this case $D(-\infty)$ is different from zero, and (\ref{d}) 
predicts an enhancement of the diffusivity at the surface compared 
to that of the bulk.
We have plotted the behavior
of $D(x)$ in Fig. \ref{f4} for different temperatures. It should be
noted that the
integrated diffusivity ${\bf D}\equiv \int D(x)dx$ tends to a finite value
${\bf D}
\sim D_0\sqrt{J/\gamma}\exp\left(\frac{-A}{T_0\alpha h_m}\right)$
when $T\rightarrow T_0^-$, 
This is at variance with what happens for melting of crystal
surfaces, where ${\bf D}\sim l$. Hence surface melting is in a
sense much weaker on the glass surface than on the 
crystal surface.

Experimental detection of the liquid layer predicted by theory
to exist at the glass-vapor interface should be possible. 
One method
might be to look for ``slumping'' (thinning of one end and thickening of
the other end) of initially shape-controlled 
glass samples in a centrifuge. 
The shapes expected for bulk-flow-induced and for surface-flow-induced 
slumping are different. 
Generally speaking bulk flow under acceleration
should preserve sharp edges and flat profiles, 
transforming {\it e.g.,} a rectangular shape to a trapezium
\cite{yvonne}. Surface flow under parallel acceleration will
modify a given profile $y(x)$ 
in time according to an equation of the type \cite{unpub}
\begin{equation}
\frac{\partial y}{\partial t}=C\frac{y''}{(1+y'^2)^{3/2}}
\label{nonlin}
\end{equation}
where $y'=\partial y/\partial x$, and 
$C$ is roughly proportional to $D$ and to acceleration $g$.
This sort of nonlinear heat conduction evolution will not generally preserve 
sharp edges and flat profiles. Given for example an initial
step function $y(x,t=0)=y_0\theta(x)$ , it will evolve in the following
manner. First, both edges, the upper one $y_0<y<y_1(t)$, and 
the lower one $y_2(t)<y<0$ will become smeared, while a central 
window  $y_2(t)<y<y_1(t)$ of the face will remain flat. As the 
spatial extension of the smeared corners increases with time,
the window will gradually shrink, and after a critical time
$t_c$ it will close, eliminating all traces
of a flat face in the slumped profile. The time $t_0$ required
for the surface smearing front to advance by $y_0$ is roughly given by
$t_0\sim y_0^2\eta_0/(l_0^3 \rho g)$, where $\eta_0$ is the viscosity of
the superficial defrozen film, i.e., a value typical of the liquid,
$\rho$ is the density, and $l_0\sim\sqrt{J/\gamma}$ is the film thickness.
The possibility of observing this effect (macroscopically, by reflectivity, 
and even better microscopically, by some surface topographic technique) seems
quite plausible.

In surface sensitive calorimetry, alternatively, a progressive 
defreezing of the surface film could
yield a detectable mark. Surface melting
of  crystals yields a singularity in the specific heat of
the form $\sim (T_M-T)^{-1-1/(\nu - 3)}$ when $T$ approaches 
$T_M$ from below, (if long-range forces decay as $1/r^\nu$).
In the glass, integrating the enthalpy
profile (\ref{hs})
across the interface we obtain a singularity in the specific heat
of the form $\sim -\varepsilon \sqrt{J/\gamma}\ln(T_0-T)$ 
when $T\rightarrow T_0^-$. 
This is again a divergence, as in crystal surface melting,
but now a much weaker one.
In compound or polar glasses, the surface static dielectric permittivity
could also be a useful tool. Microscopically, finally, liquid-like 
surface flow could be detectable by surface-specific techniques, for 
example thermal atom scattering.

Glass surfaces can also be very effectively simulated. 
A simulation study of Pd$_{80}$Si$_{20}$ published some time 
ago does provide a first qualitative suggestion for surface 
defreezing in a realistic glass forming system\cite{ballone}. The 
glass surface roughness drop demonstrated in that work upon cooling
could now be further elaborated upon in terms of suppression of capillary
fluctuations due to confinement of the liquid film. 
Our present mean-field theory 
does not include capillary fluctuations.
Their effect on the liquid film 
thickness is known to be at most
marginal in three dimensions in crystal surface melting \cite{lipowsky}, 
but will require a specific treatment in the glass case.   Experimental
evidence of surprisingly flat glass surfaces \cite{flatglass} 
do exist and might be related, although the effect of gravity 
is not discounted. This is a line
of research that will deserve further theoretical and experimental effort.

We acknowledge support from MURST COFIN97 and COFIN99, and from INFM/F,
through PRA LOTUS.


\begin{references}
\bibitem{nota} In experiments, the system freezes before the Kauzman
temperature is reached, 
due to slow dynamics. However, the form
of, for instance, the specific heat as a function of $T$ 
measured at a nonzero quenching rate is
qualitatively similar to the ideal one, while the `effective' 
glass temperature observed is in general higher
than $T_0$, and the transition rounded. The present 
analysis of surface defreezing
should qualitatively carry over to a realistic situation
where finite, albeit long, equilibration times are allowed for.
\end{references}
\end{document}